# What We Know About Using Non-Engagement Signals in Content Ranking

Tom Cunningham (Integrity Institute), Sana Pandey (UC Berkeley), Leif Sigerson (Pinterest), Jonathan Stray (UC Berkeley), Jeff Allen (Integrity Institute), Bonnie Barrilleaux (LinkedIn), Ravi Iyer (University of Southern California), Smitha Milli (Cornell Tech), Mohit Kothari (LinkedIn), Behnam Rezaei (Pinterest)

## Abstract

Many online platforms predominantly rank items by predicted user engagement. We believe that there is much unrealized potential in including non-engagement signals, which can improve outcomes both for platforms and for society as a whole. Based on a daylong workshop with experts from industry and academia, we formulate a series of propositions and document each as best we can from public evidence, including quantitative results where possible.

There is strong evidence that ranking by predicted engagement is effective in increasing user retention. However retention can be further increased by incorporating other signals, including item "quality" proxies and asking users what they want to see with "item-level" surveys. There is also evidence that "diverse engagement" is an effective quality signal. Ranking changes can alter the prevalence of self-reported experiences of various kinds (e.g. harassment) but seldom have large enough effects on attitude measures like user satisfaction, well-being, polarization etc. to be measured in typical experiments. User controls over ranking often have low usage rates, but when used they do correlate well with quality and item-level surveys. There was no strong evidence on the impact of transparency/explainability on retention. There is reason to believe that generative AI could be used to create better quality signals and enable new kinds of user controls.

# 1. Introduction

In this paper, we describe what is known about using non-engagement signals in ranking content for online platforms. Our conclusions are based on a daylong workshop with experts from industry (with hands-on experience) and academia. The purpose of this workshop was to share practical knowledge about the use of engagement and non-engagement signals in content ranking, and their tradeoffs, benefits, and limitations.

We believe that there is much unrealized potential in using non-engagement signals. These signals can improve outcomes both for platforms and for society as a whole. However, there are significant technical barriers to scaling these signals (e.g., data quantity and quality, complex tradeoffs). Many of these barriers are common across platforms, so we believe that sharing and publishing best practices will benefit many platforms and their users.

Prior to the workshop, we drafted and discussed 20 propositions based on public information and personal experience. The workshop itself consisted of 8 hours of discussing, refining, and seeking consensus on these propositions. There were 18 participants in total, representing seven major content-ranking platforms (plus former employees of one more), three universities and one institute.

Following the workshop, we drafted this paper in order to summarize the discussion during the workshop and share findings out broadly. All authors reviewed and had input on the final document, and the workshop organizers made editorial decisions.

# 2. Definitions

We organize the metrics used by platforms into the following six buckets. We provide more comprehensive definitions at the beginning of the sections corresponding to each class of metric.

| **Retention** | A user-level metric that measures the active choice to use the platform, e.g. monthly active users (MAU), daily active users (DAU), or sessions. |
|---|---|
| **Engagement** | An action taken by a user on an item, e.g. click, like, comment, retweet, upvote, downvote, dwell time, watch time. |
| **Quality metric** | A score assigned to an item, independent of the viewer, with a clear quality valence, whether positive or negative. E.g. informativeness score, toxicity score, click-gap score, nudity score, spam score, etc. |
| **Item-level survey** | A user's response to a question about a specific item, e.g. "is this item informative?" "would you like to see more like this?" "is this item worth your time?" "what star rating would you give this item?" |
| **User-level survey** | A user's response to a question about their overall experience with the platform, or about their overall attitude or opinion, not attached to a specific item of content. |
| **User control** | A setting which controls ranking or visibility of future items on the feed, e.g. "follow/subscribe", "see less like this", "block this user", "allow NSFW", "mute this term." We also include "report this item" or "report this user" as controls, although filing a report does not necessarily affect the future ranking for that user. |

It is important to distinguish between senses in which a metric can be "maximized", or used as an objective for the platform. We give a hierarchy of four levels of objectives in the table below, from topline company-level goals down to weights in an algorithm. Thus when we say "platforms maximize X" there are four different ways that can be interpreted, and the terminology introduced here allows us to be more specific about the nature of the maximization. We use "Δ" to represent metrics that are typically measured with experiments.

| Objective | Description | Typical metrics |
|---|---|---|
| Topline objective | Aggregate topline metrics used as goals for a team or the whole company. | DAU, revenue, quality (e.g. prevalence of policy-violating impressions) |
| Holdback objective | Metrics used as team goals, typically over 6-12 months, measured by experimental comparison to a holdout group that did not receive the new features. | Δimpressions, ΔDAU, Δquality |
| Launch objective | Metrics used in experiments to inform whether or not to launch the feature, typically running for 1-2 weeks. | Δclicks, Δimpressions, Δquality |
| Value function objective | Metrics used in a ranking function. Here "pClick" represents the classifier-estimated probability that a specific user clicks on a specific item. | pClick, pQuality |

# 3. Engagement

**Definition of "engagement":** An action taken by a user on an item, e.g. click, like, comment, retweet, watch, upvote, downvote. We also include continuous measures of interactions, e.g. dwell time and watch time.

### 3.1 Ranking by predicted engagement causes significantly higher time-spent and retention compared to chronological ranking.

Multiple platforms reported maintaining long-term experiments which assigned users to a chronologically-ranked feed. Those users had substantially lower time-spent and retention, with the effects remaining over months or years. It is worth noting that (1) the control group in these experiments is the full production ranking function which includes many other terms besides predicted engagement; (2) the treatment groups are typically not purely reverse-chronological ranking but also include some content-quality filters and diversity rules; (3) the measured effects were only on the viewer-side, so these experiments did not estimate effects on producers or peers.

**Public evidence:**
- Most large platforms rank their feeds primarily by predicted engagement. Many switched to using predicted engagement after using chronological or some other algorithm: Facebook ([Facebook, 2023](#)), LinkedIn ([LinkedIn, 2023](#)), Instagram in 2016 ([Instagram, 2016](#)), Twitter in 2016 ([Buzzfeed, 2016](#)), and Reddit in 2021 ([Reddit, 2021](#)).
- Facebook ran an experiment giving users a semi-chronological feed. They found user time-spent declined by 3% after 10 days and was continuing to decline when the experiment ended. ([FBArchive, 2018](#))
- In 2020 Facebook and Instagram ran experiments with semi-chronological feeds: time-spent declined by 20% on FB and 10% on Instagram on average over the following 3 months. ([A. M. Guess et al., 2023](#))
- A 2022 paper reported that users in Twitter's long-term chronological holdback, which began in 2016, had approximately 38% fewer impressions/day. ([Bandy and Lazovich, 2022](#), table 2 and figure 4)

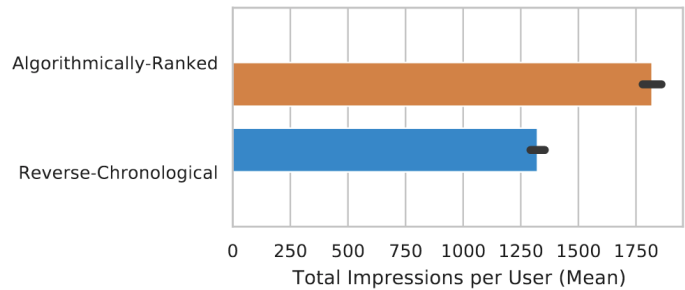

([source](#))

**3.2 The relative effects of value-model weights on retention are difficult to estimate.**

---

The core ranking algorithm is typically a linear combination of engagement predictions, e.g. weights on pClick, pFav, etc. The full ranking algorithm often has hundreds of additional terms or modifications of weights depending on contextual factors. Platforms often wish to estimate the set of weights that would maximize long-term retention; however, this is a challenging statistical problem and there is no consensus on the best practice.

Estimates of the weights that maximize retention can be derived either from experimental data or from observational data, but both pose significant challenges. Estimating from experiments requires many large and long-running experiments so, in practice, it is only practical to explore a small subset of the space of all possible weights ([Vengerov, 2023](#)). Estimating weights from observational data yields more precise estimates, but they will be contaminated by bias from unobserved differences between users and sessions that is difficult to credibly remove.

**3.3 Active engagements tend to be more valuable for user retention.**

---

Workshop participants brought up a variety of observations and generalizations about which engagements are most valuable for long-term retention (precisely: the relative size of the retention-maximizing value-model weights). A common theme was that "active" engagements tend to be more valuable:

- Longer engagements tend to be more valuable, e.g. data on how long a user spent on an item (linger-time or watch-time) is more valuable than binary data on whether a user saw an item.
- High-effort engagements tend to be more valuable, e.g. a longer comment compared to a shorter comment, a repost with text vs a repost without text, a reaction vs a like.
- High-intent engagements tend to be more valuable, e.g. bookmarking an item, following the producer of an item, registering for an event.

**Public evidence:**
- In 2012 YouTube switched from maximizing clicks to maximizing a combination of clicks and watch-time. They reported a short-term drop in clicks but a long-term increase in retention ([Goodrow, 2021](#)).

## 4. Quality Measures

**Definition of "quality":** A score assigned to an item independent of the viewer with a clear quality valence, i.e. where a high score (or low score) is more desirable in some sense, e.g. informativeness score, toxicity score, click-gap score, nudity score, spam score. Under the heading of quality, we also include scores that reflect the provenance or history of an item rather than the intrinsic quality of the content itself (e.g. scores that identify "inauthentic", "scraped", or "unoriginal" content as being low quality). Sometimes these quality scores are assigned by paid human raters, sometimes they are assigned by a classifier trained on human-labeled data, and sometimes they are calculated using hard-coded heuristics. Note that "quality" excludes item surveys, discussed below, unless the scores are aggregated at the item level.

**4.1 A positive ranking-weight on quality metrics often significantly increases long-term retention.**

Most platforms put substantial ranking weight on quality metrics, implying they sacrifice user engagement (at least in the short run) to increase the quality of items shown. There are many reasons why platforms may care about quality.[1] Participants generally believed that many types of quality metrics significantly increase long-term user retention.

**Public evidence:**
- Many platforms have publicly described quality metrics they use in ranking:
  - Google search uses search quality guidelines (a paid rater program) to label content ([Google, 2023](#)).
  - YouTube uses predicted labels for "sensationalistic tabloid content" and "authoritative content" ([Goodrow, 2021](#)).
  - LinkedIn predicts whether posts share "knowledge and advice" and uses that prediction in ranking ([Entrepreneur, 2023](#)).
  - Facebook uses professional rater scores (scores by "inform, connect, entertain"), broad trust scores (measuring the diversity of engagement) ([Wired, 2018](#)), third-party misinformation labels, click gap (the volume of organic traffic a website receives) ([CNBC, 2019](#)), search gap (the volume of search traffic a page receives) ([FBarchive, 2022](#)), network centrality score (the "pagerank" score of a website) ([CNBC, 2019](#)).
- Facebook assigned a random group of users to have fewer quality terms in their ranking function ("minimal integrity holdout"). The biggest impact was on disabling the "clickbait" and "adsfarm" quality terms. After 1 month these users had higher overall activity by most metrics, e.g. impressions was up by around 0.4% ([FBArchive, 2019](#)). However, after 2 years these users had lower overall activity ([Horwitz, 2023](#), p311).
- YouTube reduced the ranking score of videos classified as "trashy" or "tabloid-style." After three weeks watch time was down by 0.5%, but after three months watch time recovered and increased relative to a holdout group. (Goodrow quoted in [Doerr 2017](#), p. 164).

---

[1] Some examples of other reasons why platforms care about "quality" metrics, outside of user retention: to satisfy advertisers, regulators, app stores, the media, and for intrinsic ethical reasons.

**4.2 Predicted engagement often has a negative relationship with quality scores.**

Many participants reported finding negative correlations between various quality scores and predicted engagement (or actual engagement). These negative relationships do not occur everywhere but many participants said they were quantitatively large and constituted important problems in ranking.

Note that it can be difficult to infer these correlations using only data on the subset of items that are shown to a user when feeds are ranked using both engagement predictions and quality scores. As a consequence, the correlation between quality and predicted engagement cannot be transparently inferred from either (1) comparing the average quality of impressions seen in a ranked vs a chronological feed, or (2) calculating the correlation between engagement and quality among the subset of items seen by users with a ranked feed. Nevertheless, participants felt that there was often sufficient evidence to infer an underlying negative relationship between quality and engagement.

**Public evidence:**
- Mark Zuckerberg wrote, "as a piece of content gets close to that line [of what is allowed], people will engage with it more on average – even when they tell us afterwards they don't like the content." (Zuckerberg, 2018)
- Facebook found in 2019 that a substantial share of the most popular identity-based groups and pages in the US (African American, Christian, Native American) were run by non-American administrators, primarily from Kosovo and Macedonia, whose main tactic was reposting copied content that had high engagement rates. (Jeff Allen, 2019)
- In 2019, an internal Facebook analysis also found that items with higher predicted engagement rates had significantly higher likelihood of being classified as low quality using Facebook's own internal "Feed Unified Scoring System" (FUSS). (Jeff Allen, 2019)
- An internal Facebook analysis also found that "for posts with very high reach (especially the top 1-2 percentile) GFTW prevalence is lower and BFTW prevalence is higher than average." Here GFTW and BFTW are averages over user responses to survey questions about whether these posts are "good for the world" or "bad for the world." (FBArchive, 2020)
- A meta-analysis from 2021 found that on Twitter, "each message is 12% more likely to be shared for each additional moral-emotional word." (Brady and Van Bavel, 2021).
- A study of Facebook and Twitter posts from 2021 found that "posts about the political out-group were shared or retweeted about twice as often as posts about the in-group" and "the average effect size of out-group language was about 4.8 times as strong as that of negative affect language and about 6.7 times as strong as that of moral-emotional language." (Rathje, Van Bavel, and van der Linden, 2021)
- Over 2021 and 2022 Facebook released dataset of their 20 most-viewed links and posts. Between 60% and 80% of these posts failed some basic quality checks, either "the account behind it is anonymous, is posting unoriginal content, using spammy page or group networks, or the post or link violated Facebook's community standards." (Integrity Institute, 2022)
- An analysis of Mexican front-page news articles in 2021 on Facebook found that a two-SD increase in "angry," "sad," "love" and "wow" reactions correlated with a 10-21% percent increase in reshares, with "angry" having the strongest effect of 21%. (De León and Trilling, 2021)
- An analysis of 2022 tweets from US Senators compared many types of language and found that "negative emotion", "political outgroup", "moral outrage," "power," and "greed" words had the strongest positive correlations with shares and favorites (Mercadante, Tracy, and Götz, 2023)
- A number of participants recounted negative relationships between quality and "sharing" engagements, e.g. reposting, retweeting, resharing. The relationship was especially strong for items that had relatively "deep" share chains, or had high predicted "downstream" engagement.
- A number of participants reported finding that negative relationships between quality and engagement were stronger within content that is political or health-related:

- Half of Facebook's most-engaged-with posts related to the 2016 US election were misinformation. This far exceeds the average rate of misinformation, i.e. the most-engaged-with posts have a much lower-than-average quality. ([Buzzfeed, 2016](#))
- Multiple leaked analyses from Facebook showed that highly-engaging civic and health content has low quality. ([Wall Street Journal, 2021](#)) ([Wall Street Journal, 2023](#))
- In 2021, Facebook adjusted engagement weights for ranking political ("civic") content: "We take away all weight by which we uprank a post based on our prediction that someone will comment on it or share it" ([Wall Street Journal, 2023](#)). Two other platforms attending our workshop also reported using modified ranking for civic and health content.
- In a survey-based 2022 experiment where people were shown tweets on their personalized (predicted engagement-ranked) timeline and asked, for each one, whether they wanted to see tweets of that type, their answers were 0.18 SD lower for political tweets. [(Milli et al. 2023)](#)
- Two studies found that chronological ranking had higher shares of low-quality content than engagement-based ranking, in contrast to the pattern from other studies. However, note that in both of these cases, the engagement-based ranking also was modified to incorporate quality proxies.
  - Facebook and Instagram found that assigning users to chronological ranking in 2020 increased the share of "untrustworthy" impressions by 70% on FB and by 20% on Instagram. ([Guess et al., 2023](#))
  - A 2022 paper reported that users in Twitter's long-term chronological holdback had approximately 8% higher rate of "marginally abusive" impressions per day. ([Bandy, 2022](#), p5).

**4.3 Quality is often positively related to the diversity of engagement across viewers.**

A number of platforms have built metrics of the "diversity" of the users who engaged with an item, and there was a common perception that diversity tended to be positively correlated with quality scores. This approach is sometimes called "bridging-based ranking." [(Ovadya and Thorburn, 2023)](#)

**Public evidence:**
- A 2020 internal Facebook study on ranking comments found that "content that receives engagement from a [politically] diverse audience tends to have much higher quality compared to content that a [politically] homogenous audience engages with." Indicators of quality included reduced engagement with bullying comments and more "affiliative, helpful, and respectful language." ([FBArchive, 2020](#))
- Twitter's "birdwatch" algorithm (AKA "community notes") prioritizes notes with more-diverse engagement, where diversity is defined as the axis that most divides people over a particular tweet. In 2022, Twitter's researchers found that such notes scored higher both in terms of how helpful they are rated and the correction of false information. ([Wojcik et. al., 2019](#))

# 5. Item-Level Surveys

**Definition of "item-level survey":** a user's response to a question about a specific item. Examples include "Is this post worth your time?" ([Instagram, 2023](#), [Facebook, 2021](#)), "Do you want to see more or less posts like this in Feed?" ([Facebook, 2022](#)), "Was the comment above high or low quality?" ([Social Media Today, 2018](#)), "What did you think of this video?" ([Goodrow, 2021](#)), or asking if an item is "informative", "inspiring", "funny", "misleading" etc. (Note that the phrases "see more" and "see less" have been used both in surveys and in user controls).

Unlike "engagement" with an item or use of "controls" at the item level, a survey is typically a random sample; it is only displayed for a small fraction of all users and impressions. While such data is sparse, this randomness can be an advantage. It is common for a small fraction of users and items to account for the majority of engagements, and some users use controls much more than others. Thus both user controls and engagement suffer from selection bias, and both can also be gamed.

Surveys can provide information that engagement signals do not. For example, if someone "likes" a picture of a celebrity at the Grammys, do they like the Grammys, the artist, or the dress she's wearing?

YouTube runs a survey asking about positive qualities of videos. Here's an example: ([Web Archive, 2019](#)).

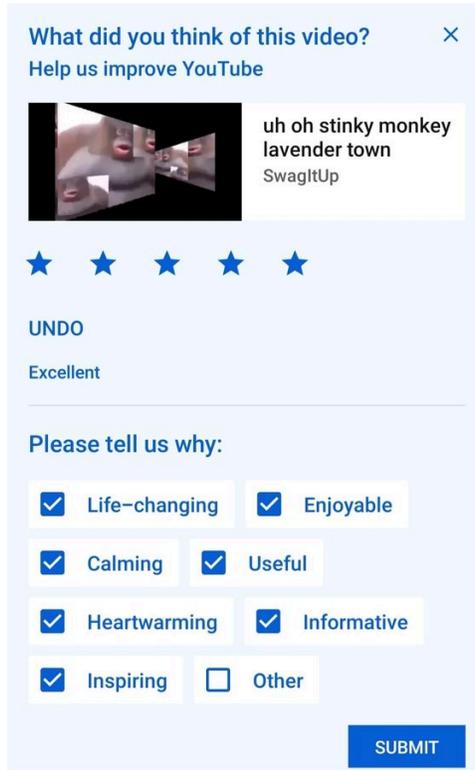

Facebook implemented "see more/see less" surveys on items in 2022: ([Facebook, 2022](#))

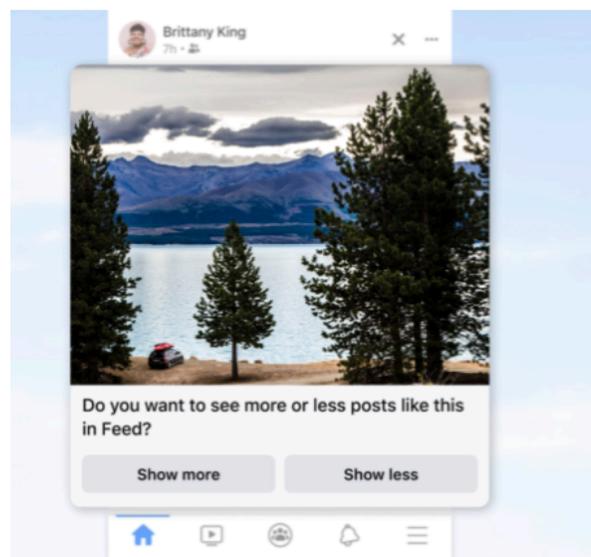

## 5.1 Multiple platforms have built predictive models for item-level survey responses.

Multiple platforms have built predictive models for item-level survey responses using a variety of features, including content-level, user-level, and engagement-related features. This is no different in principle from predicting engagement (whether a user will like, reply, watch, etc.) In our workshop, five of seven platforms indicated that they were already using such surveys.

One platform reported that such predictive models achieve a typical correlation coefficient of around 0.2 with actual survey responses, though better is possible in certain cases.

Several platforms noted that the same survey prediction model performed differently on different types of content, e.g. cats vs politics, or for different users, so it might be necessary to target additional survey data collection in order to build a more broadly accurate model.

**Public evidence:**
- A 2019 paper documents how YouTube incorporates "satisfaction objectives" including survey data, predicting survey responses using a deep neural net and using these predictions in ranking. (Zhao et al., 2019)
- YouTube discusses the use of surveys and says "based on the responses we do get, we've trained a machine learning model to predict potential survey responses for everyone." (Goodrow, 2021)
- Facebook has written about their use of models predicting "worth your time," "inspirational," and other item-level surveys. (Facebook, 2021)
- Instagram says "We survey people and ask whether they find a particular reel entertaining or funny" and that they predict these outcomes. (Instagram, 2021)
- Instagram recently wrote about surveys as one of four classes of ranking signals, mentioned a number of item surveys used by different platforms, and noted that predictive models are necessary: "To scale survey results, platforms need to overcome challenges of extrapolation." (Arcamona et. al., 2023)
- A 2019 model predicting survey answers to "was the comment above high or low quality?" is reported in Facebook documents (FBArchive, 2019). These surveys correlated with lower measures of bullying/incivility, higher likelihood of a love reaction, and models using these surveys reportedly "performed well" across non-English languages (FBArchive, 2019).
- Correlation of answers to "was the comment above high or low quality?" with a predictive model, from a 2019 Facebook survey across 20 countries (FBArchive, 2019):

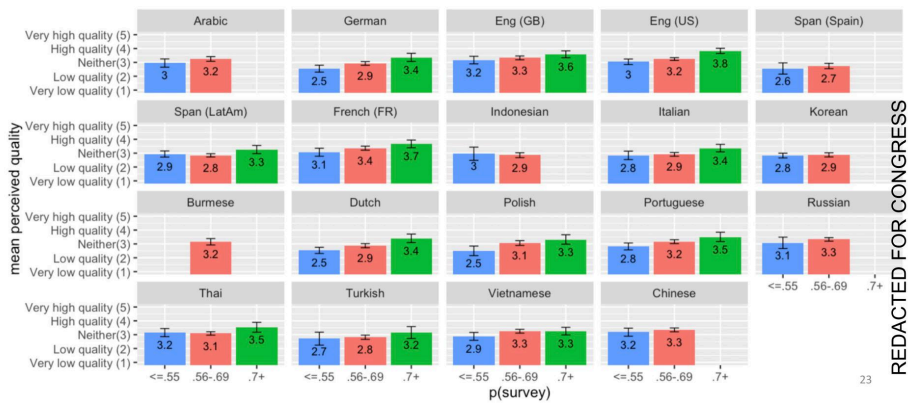

## 5.2 Responses to item-level surveys are sensitive to wording.

It is well known from the survey literature that responses can be sensitive to wording, and the context of recommenders is no different. However, there are specific issues that apply here.

Several participants noted that they have often found significant differences between "Does this post make you angry?" and "Do you think this post would make people angry?" or "Do you like this post?" and "Is this post good for the world?" There is a long history of surveying people in this indirect way to avoid certain types of response bias ([Kwak et al., 2019](#)).

Several participants mentioned that it is hard to get answers other than 4 or 5 on five-point scales. For example, "worth your time" is a low bar, and in practice does not effectively differentiate between "somewhat" and "very" worth your time. This explains why some surveys have had mostly superlative options, for example a YouTube survey ([Reddit, 2018](#)) with options "not good," "good," "very good," "extremely good," and "absolutely outstanding." ([Google, 2019](#))

## 5.3 Item-level survey responses are typically positively correlated with item quality.

Multiple platforms reported experimenting with survey wording to find questions that correlated with pre-existing quality (or "integrity") metrics and outcomes.

Surveys can be used to increase the legitimacy of certain types of content demotion, by letting users indicate which types of content they consider low quality – or proving that pre-existing quality measures are correct. It is possible to experiment with survey wording to find an innocuous question that justifies a more controversial intervention. Elsewhere, former Facebook employees have claimed that the company chose the item-level "good for the world" survey question because it correlated with other measures of misinformation. ([Cunningham, 2023](#))

**Public evidence:**
- Including predicted user responses to a "worth your time" survey when ranking political content reduced views of misinformation by 12% during a 15-day test at Facebook in 2020 ([FBArchive, 2020](#)).
- Facebook asked "Is this kind of post good for the world?" in 2020 (Pahwa 2021). Per internal documents, "the question is meant to support demotion models that would reduce unwanted negative experiences." ([FBArchive, 2020](#))

## 5.4 There is evidence that using item-level survey responses in ranking helps retention.

Multiple platforms reported that at least one of their item-level survey prediction models correlated with long-term retention. However, they noted that not all surveys have done so and that this depends both on the survey wording and the definition of retention (e.g. views vs. sessions, and over different time scales.) We do not know of any clear public evidence on this point.

# 6. User-Level Surveys

**Definition of "user-level survey":** a user's response to a question about their overall experience or attitude, not about a specific item of content. We include surveys about experience over a fixed time period (e.g. the last session, the last day, the last week).

User-level surveys are fairly common. The "rate your YouTube experience today" survey is an example, and other platforms reported working with session-level surveys (where the user is asked about their last session at the beginning of the next session.) User-level surveys also include questions about on-platform negative experiences of all kinds, general measures of user satisfaction, and a wide variety of measures of well-being, polarization, etc.

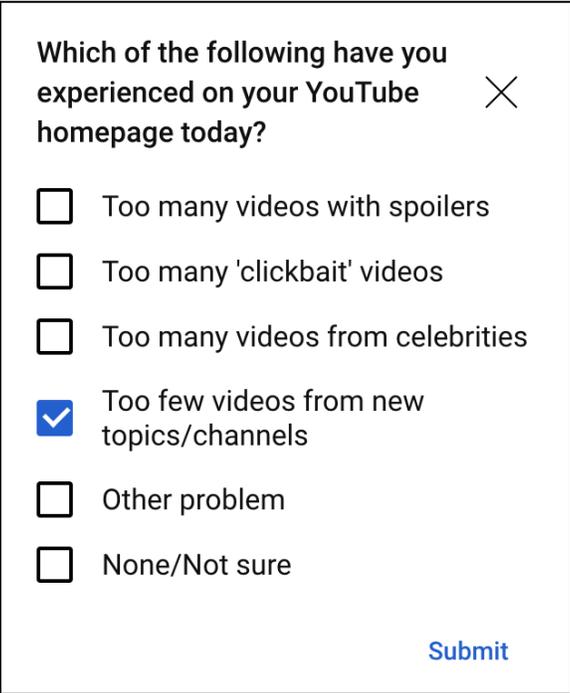

A session-level survey from YouTube, 2023

**6.1 Ranking changes often significantly affect self-reports of content exposure and experiences.**

Self-report exposure measures ask about seeing specific content. They can sometimes be moved mechanically, for example, a question like "Have you seen posts with nudity?" can be moved by changing the ranking weight on a nudity classifier. Other experiences are more complex, e.g. "Have you experienced bullying?" but there are reports of moving this survey result too.

User-level surveys can detect unwanted experiences that other measures do not. While many platforms have classifiers for various types of harmful content, e.g. harassment as decided by third-party labelers, these classifiers do not necessarily capture the experience of harassment. This is a result of several factors: classifier inaccuracy, differences in definition between platforms and users, and the contextual factors that shape how the content a user sees translates into user experiences (for example, the relationship between the user and the harasser, or pre-existing stresses in a user's life).

**Public evidence:**
- Meta ran an ongoing user-level survey called TRIPS for "tracking reach of integrity problems survey" across Facebook and Instagram, tracking what percentage of users reported a variety of experiences over the last seven days, including bullying and harassment, nudity, inflammatory material, misinformation, links to offsite pages with too many ads, impersonation, etc. ([FBArchive, 2020](#)).

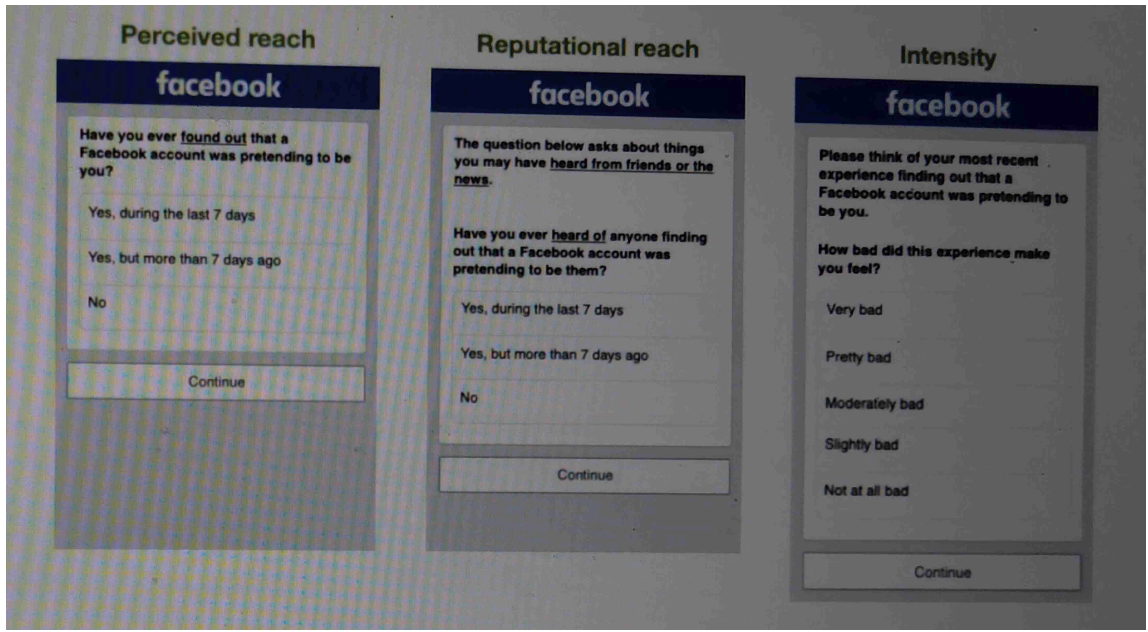
Three kinds of measures of user-reported experience in Meta's TRIPS survey program, according to a 2020 document (FBArchive, 2020)

- In one 2020 experiment, Instagram found that recent product changes had reduced not the prevalence but the reported severity of bullying, as compared to a holdout group (FBArchive, 2020).
- Platforms have used comparative metrics of user experience to gauge performance within products and against competitors (FBArchive, 2022).  Many of these results (e.g. perceived negative experiences with political content) have been replicated externally (USC Neely Center, 2023).

**6.2 Ranking effects on self-reported satisfaction with the platform are difficult to move.**

Two platforms reported using user-level product satisfaction surveys to evaluate product launches (via experiments), or to evaluate long-term progress (via holdbacks). There was agreement that most ranking changes did not cause big enough movements in such measures to be reliably detected, although occasionally there were larger effects. The large sample sizes required to detect small effects means that only the largest platforms are likely to be able to move these metrics in experiments.

**Public evidence:**
- In a 2018 holdback test, Facebook's implementation of MSI-based ranking increased the number of people reporting the highest level of user satisfaction by 0.5%. (FBArchive, 2018)
- Large changes, like Facebook's shift to MSI as an objective function, measured across large samples produced non-significant or barely significant results for measures of newsfeed satisfaction, relevance and whether newsfeed showed people important to users. (FB Archive, 2019)
- Effects of ranking changes on whether Facebook "Cares about You" or whether users are satisfied with Facebook products have generally not been statistically significant, even with significant product changes and large sample sizes. Per one document, sentiment metrics are "often too noisy to give guidance in everyday product decisions." (FB Archive, 2020)

**6.3 Ranking effects on broad outcomes, such as well-being, polarization etc. are typically below 0.05sd after 1-6 months.**

Broad measures such as general well-being, life satisfaction, polarization, or attitudes toward the company usually do not show statistically significant changes due to ranking even with relatively large and long-running experiments.

However, extrapolating these results to overall individual and societal effects is both methodologically and definitionally complex, so null results in experiments reported so far cannot be straightforwardly interpreted as meaning platforms have no effects on these broader outcomes. For example, many content publishers report being sensitive to ranking effects in ways that are not captured by individual-level ranking experiments.

As one platform put it, "a narrow scope works." It is possible to reduce the experience of unwanted contact from strangers for teens, but quite difficult to increase general well-being. Another platform noted that *sometimes* changes in broad measures are possible. They suggested experimenting with large ranking changes, and that session-level surveys are a useful intermediate measure between item-level surveys and surveys measuring broad attitudes or outcomes.

**Public evidence:**
- Significant changes to ranking have typically moved such broad measures by less than 0.05sd over six weeks, implying large sample sizes are needed to detect them (Guess et al. 2023a, 2023b; Nyhan et al. 2023)
- A 2018 Facebook survey of 100,000 people evaluating recent "meaningful social interaction" ranking changes versus a six month holdout group found no difference in general well-being (FBArchive, 2018).
- An internal Instagram study found that hiding likes reduced the reported incidence of negative social comparison by 2% but had no effect on general well-being measures (Meta, 2021; Wells, Horwitz and Seetharaman 2021)

**6.4 There are sometimes significant effects for specific subgroups.**

It may be that case that e.g. a 1% change for all users is actually a 10% change for 10% of users.

Several platforms reported that they frequently see large differences in response to ranking changes for people in different countries, of different ages, and who use the product often vs. rarely. Others felt that the majority of ranking changes had fairly stable proportional effects on most metrics across most groups, and quantitatively important heterogeneities were relatively rare.

Proportional effects will affect users differently in absolute terms if there is wide variation in current baselines. For example, older and more conservative people may be exposed to more misinformation in their feeds (Gonzalez-Bailon et al., 2023; Guess, Nagler, and Tucker, 2019)

There may also be outsized effects on specifically vulnerable subgroups. Consider that diet videos may be fine for many people, but someone with an eating disorder could be disproportionately affected by a feed with many diet videos in it. Correspondingly, there is work on optimizing ranking for subgroups. (Singh et al., 2021) describe a change to YouTube's recommender to reduce the amount of "unhealthy" content viewed by the top 5% most exposed users (as opposed to the average exposure).

# 7. User Controls

**Definition of "user control":** User controls are typically accessed either (1) attached to an item in a ranking surface, or (2) in a "settings" surface. These give the user direct control over what is ranked, e.g. "see more/see less", "block this user", "block this topic", "filter NSFW", explicit topic selection, choice of chronological vs ranked.

As opposed to surveys, user controls are always available in the UI. Although reaction buttons such as "like" or "favorite" might be considered controls, we call them engagement instead because (a) controls directly affect what is shown in feeds and (b) such reactions are social, that is, visible to other users.

**7.1 For most platforms, user ranking controls have low uptake, with a few percent of users at most.**

Most platforms indicated that the portion of their user base actively utilizing user controls outside of follow controls remains within the single-percentage range. Some platforms have used various techniques to increase participation rates, such as (1) onboarding for settings/controls, (2) prompting users to use controls at session points, or (3) improving the discoverability of the control by exposing it on each item. One platform has shown that this can improve participation by as much as a factor of 10.

Participants noted that, while user controls have low usage rates, they are sometimes useful signals. Opting in or out of topics such as explicit content may instead function as a stronger indicator of user value or engagement.

**Public evidence:**
- A 2021 internal Meta document, quoted in a 2023 court case, found that "when [D]aisy controls are opt-in, only 0.72% of people choose to hide like counts, but when they're opt-out, 35% leave their like counts hidden." [(Commonwealth of Massachusetts v. Meta Platforms, 2023)](#)
- The same case referenced 2022 Meta internal documents that had found young users rarely change their settings, with surveys reporting it "overwhelming to try and change their notification settings" [(Commonwealth of Massachusetts v. Meta Platforms, 2023)](#).
- Research conducted on Twitter in 2022 concluded that the "percent of all users that opt out [of the algorithmic feed] is between one and ten percent." [(Milli, Belli, and Hardt, 2022)](#)

**7.2 Negative controls are often highly correlated with quality and item level surveys.**

Platforms have found that sparse negative feedback (ie: "x"-out, downvoting) can help identify low-quality content in the ranking process. These signals can be as valuable as social engagement in predicting quality or item-level survey outputs.

User controls are often used strategically to influence ranking for other users, e.g. strategically reporting or blocking content. Some platforms have taken actions to mitigate this vulnerability by capping or normalizing signals from controls. One potential avenue platforms have explored is generating trust scores on a user level, which can be used to determine if content should be immediately downranked or sent for human review. These scores can also be used to identify brigaders (actors artificially increasing or decreasing ranking through coordinated efforts).

**Public evidence:**
- A 2019 study found evidence of users actively harnessing user controls on Twitter (i.e. reporting tweets or accounts for site policy violations, x-ing out posts) to "benefit themselves, their Twitter account…or a particular topic or cause." [(Burrell, Kahn, et. al., 2019)](#)
- A Facebook analysis in 2022 found that negative reactions only get ~0.5% of the usage that positive reactions get ([FB Papers, 2022](#), page 3). However, incorporating "lightweight negative feedback" (such as the X in the upper right corner of each post) into a "p(dislike) model" used in ranking increased several item-level survey measures compared to control, including "worth your time" (+3.3%), "helps to understand political issues" (+3.9%), and "leads to useful political discussions" (+4.3%). It also reduced the prevalence of "viral inflammatory content" by 8-10%. ([FB Papers, 2022](#))
- Facebook also found that diverse negative reactions to content aggregated at the actor level successfully identify low-quality content, shifting metrics such as "is this high quality" by more than 2% compared to control conditions. ([FB Papers, 2022](#))
- LinkedIn revealed in 2023 that user reports are used to dynamically rank a content queue for human review, to prioritize limited human reviewer bandwidth ([Chandak and Verma, 2023](#))

# 8. Transparency

**8.1 There was no strong evidence on the impact of transparency/explainability on retention**

While there is a widespread view that some form of transparency or explainability is worthwhile, there are very few examples of the effect of transparency or explainability on retention measures.

Many participants highlighted the distinction between transparency and explainability within recommender systems. While transparency operates on an algorithm-wide level, explainability is tailored to personalized recommendations on a user basis and can often be integrated as a separate step in the ranking process.

One platform reported that integrating "person X also clicked Y" explanations into the user interface caused increased user retention, engagement, and the exploration of new content. They attributed this to the social nature of the explanation.

**Public evidence:**
- A 2021 internal Facebook review noted the generally positive reception of "Why Am I Seeing This" by media and the public. ([FBArchive, 2021](#), slide 118332)

# 9. Applications for Generative Artificial Intelligence

We allocated one hour during the workshop for topics chosen by the attendees. Generative AI was selected as one of these topics. Our discussion in this area was understandably much more speculative than in other topics.

**9.1 Generative AI may help scale quality measures.**

Some participants agreed that generative AI may help measure content quality, especially in areas where humans are traditionally needed to manually label content (e.g., policy violations). There was consensus that generative AI is likely to provide a cheaper, faster option, which may allow more comprehensive content monitoring. Though participants agreed on the potential impact of Gen AI in this area, there were significant open questions on how this could be done, such as the need to keep human involvement in individual content moderation decisions.

**9.2 Generative AI may unlock more user control and explicit feedback.**

Some participants agreed that a generative AI "assistant" in content rankers could help solve the low adoption rates seen in user controls and other forms of explicit feedback. Though users often want control of what content they see, their needs are too varied to be captured by a static form of user input. Generative AI could help solve this problem by providing a flexible natural-language input for users.
However, participants also noted the many barriers to scaling this functionality, including: a) getting users to write full sentences to the assistant (given very low adoption rates for user controls with simpler input) and b) giving the GenAI assistant sufficient information about content.

# 10. Conclusions

This workshop was a unique opportunity to collect and document industry best practices, and the evidence behind them. Although much of this practice remains private, we were at least able to establish, gauge consensus for, and document the public evidence for a set of propositions which seem to apply generally across many types of platforms. This is possible in part because companies have been more forthcoming about some of these issues in recent years, and in part because of recent leaks.

In the opinion of the authors, the most promising near-term strategies are more widespread use of item-level surveys and better item quality proxies.

Beyond that, there are a number of propositions for which there is little public information – either little has been published, or the relevant experiments have not been performed. This includes the use of user-level surveys for steering content selection towards different kinds of long term outcomes, including well-being, polarization, etc., and the effects of transparency and explanations on retention.

# Acknowledgments

We would like to thank Spencer Gurley for his support in preparing the bibliography, as well as the UC Berkeley Center for Human-Compatible AI, USC Marshall School's Neely Center and the Integrity Institute for helping us make the workshop possible.